\documentclass[12pt,epsfig]{article}
\usepackage{graphicx}
\usepackage{array}
\usepackage{amsmath,amssymb}
\usepackage[latin1]{inputenc}

\begin{document}
\begin{titlepage}

\title{Lorentz and $SU(3)$ groups
 \\ derived from cubic quark algebra}
 \author{Richard Kerner$^{\star}$ }

\maketitle



\begin{abstract}
{\small We show that the Lorentz and the $SU(3)$ groups can be
derived from the covariance principle conserving a $Z_3$-graded three-form
on a $Z_3$-graded cubic algebra representing quarks endowed with non-standard 
commutation laws. }
\end{abstract}

\footnote{ \small $^{\star}$
  Laboratoire de Physique Th\'eorique de la Mati\`ere Condens\'ee, \\\small
  Universit\'e Pierre-et-Marie-Curie - CNRS UMR 7600 \\\small
  Tour 22, 4-\`eme \'etage, Bo\^{i}te 121, \\\small
  4, Place Jussieu, 75005 Paris, France\\\small
  \small}
\end{titlepage}

\newpage

1. Many fundamental properties of matter at the quantum level can be 
announced without mentioning the space-time realm. The Pauli exclusion principle, 
symmetry between particles and anti-particles, electric charge and baryonic number
conservation belong to this category. Quantum mechanics itself can be formulated 
without any mention of space, as was shown by Born, Jordan and Heisenberg \cite{BornJH} in their
version of matrix mechanics, or in J. von Neumann's \cite{JvNeumann} formulation of quantum theory in terms
of the $C^*$ algebras. The non-commutative geometry \cite{MDVRKJM} gives another example of interpreting
the space-time relationships in pure algebraic terms. 

Einstein's dream was to be able to derive the properties of matter, and perhaps its very existence,
from the singularities of fields defined on the space-time, and if possible, from the
geometry and topology of the space-time itself. A follower of Maxwell and Faraday, he
believed in the primary role of fields and tried to derive the equations of motion
as characteristic behavior of field singularities, or the singularities of the
space-time (see e.g. \cite{EinsteinInfeld}).

But one can defend an alternative point of view supposing that the existence of matter
is primary with respect to that of the space-time. In this light, the idea to derive the geometric 
properties of space-time, and perhaps its very existence, from fundamental symmetries and interactions
proper to matter's most fundamental building blocks seems quite natural. 

If the the space-time is to be derived from the interactions of
fundamental constituents of matter, then it seems reasonable to choose the strongest ineractions
available, which are the interactions between quarks. The difficulty resides in the
fact that we should define these ``quarks" (or their states) without any mention of space-time.

The minimal requirements for the definition of quarks at the initial stage of model building
are the following:

\indent
\hskip 0.5cm
{\it i} ) The mathematical entities representing quarks should form a linear space over complex numbers, 
so that we could produce their linear combinations with complex coefficients.

\indent
\hskip 0.5cm
{\it ii} ) They should also form an associative algebra, so that we could consider their multilinear 
combinations;

\indent
\hskip 0.5cm
{\it iii} ) There should exist two isomorphic algebras of this type corresponding to quarks
and anti-quarks, and the conjugation transformation that maps one of these algebras onto
another, ${\cal{A}} \rightarrow {\bar{\cal{A}}}$.

\indent
\hskip 0.5cm
{\it iv} ) The three quark (or three anti-quark) and the quark-anti-quark combinations 
should be distinguished in a certain way, for example, they should form a subalgebra
in the algebra spanned by the generators. 

With this in mind we can start to explore the algebraic properties of quarks that would lead
to more general symmetries, that of space and time, appearing as a consequence of covariance
requirements imposed on the discrete relations between the generators of the quark algebra.

2. At present, the most successful theoretical descriptions of fundamental interactions are based
on the quark model, despite the fact that isolated quarks cannot be observed. The only experimentally
accessible states are either three-quark or three-anti-quark combinations (fermions) or the quark-anti-quark
states (bosons). Whenever one has to do with a tri-linear combination of fields (or operators),
one must investigate the behavior of such states under permutations.  

Let us introduce $N$ generators spanning a linear space over complex numbers,
satisfying the following relations which are a cubic generalization of anti-commutation in
the ususal (binary) case (see e.g. \cite{Kerner3}, \cite{VARKBLR}):
\begin{equation}
\theta^A \theta^B \theta^C = j \, \theta^B \theta^C \theta^A = j^2 \, \theta^C \theta^A \theta^B,
\label{ternary1}
\end{equation}
with $j = e^{i \pi/3}$, the primitive cubic root of $1$. We have ${\bar{j}} = j^2$ and $1+j+j^2 = 0$.
We shall also introduce a similar set of {\it conjugate} generators, ${\bar{\theta}}^{\dot{A}}$,
$\dot{A}, \dot{B},... = 1,2,...,N$, satisfying similar condition with $j^2$ replacing $j$:
\begin{equation}
{\bar{\theta}}^{\dot{A}} {\bar{\theta}}^{\dot{B}} {\bar{\theta}}^{\dot{C}} = 
j^2 \, {\bar{\theta}}^{\dot{B}} {\bar{\theta}}^{\dot{C}} {\bar{\theta}}^{\dot{A}} 
= j \, {\bar{\theta}}^{\dot{C}} {\bar{\theta}}^{\dot{A}} {\bar{\theta}}^{\dot{B}},
\label{ternary2}
\end{equation}
Let us denote this algebra by ${\bf{\cal{A}}}$.
We shall endow this algebra with a natural $Z_3$ grading, considering the generators $\theta^A$
as grade $1$ elements, and their conjugates ${\bar{\theta}}^{\dot{A}}$ being of grade $2$.
The grades add up modulo $3$, so that the products $\theta^{A} \theta^{B}$ span a linear
subspace of grade $2$, and the cubic products $ \theta^A \theta^B \theta^C$ are of grade $0$.
Similarly, all quadratic expressions in conjugate generators, ${\bar{\theta}}^{\dot{A}} {\bar{\theta}}^{\dot{B}}$
are of  grade $2 + 2 = 4_{mod \, 3} = 1$, whereas their cubic products are again of grade $0$, like
the cubic products od $\theta^A$'s.

Combined with the associativity, these cubic relations impose finite dimension on the
algebra generated by the $Z_3$ graded generators. As a matter of fact, cubic expressions are the
highest order that does not vanish identically. The proof is immediate:
\vskip 0.2cm
\centerline{$\theta^A \theta^B \theta^C \theta^D = j \, \theta^B \theta^C \theta^A \theta^D =
j^2 \, \theta^B \theta^A \theta^D \theta^C = $}
\vskip 0.2cm
\centerline{$ =j^3 \, \theta^A \theta^D \theta^B \theta^C =
j^4 \, \theta^A \theta^B \theta^C \theta^D,$}
\vskip 0.2cm
and because $j^4 = j \neq 1$, the only solution is 
\begin{equation}
\theta^A \theta^B \theta^C \theta^D = 0.
\label{quartic2}
\end{equation}
Therefore the total dimension of the algebra defined via the cubic relations (\ref{ternary1})
is equal to $N + N^2 + (N^3 - N)/3$: the $N$ generators of grade $1$, the $N^2$ independent
products of two generators, and $(N^3-N)/3$ independent cubic expressions, because the cube
of any generator must be zero, and the remaining $N^3-N$ ternary
products are divided by $3$, by virtue of the constitutive relations (\ref{ternary1}).

The conjugate generators ${\bar{\theta}}^{\dot{B}}$ span an algebra ${\bf{\bar{\cal{A}}}}$ isomorphic 
with ${\bf{\cal{A}}}$. Both algebras split quite naturally into sums of linear subspaces with definite grades:
$${\bf{\cal{A}}}= {\bf{\cal{A}}}_0 \oplus {\bf{\cal{A}}}_1 \oplus {\bf{\cal{A}}}_2, \, \ \ \, \ \, 
{\bf{\bar{\cal{A}}}}= {\bf{\bar{\cal{A}}}}_0 \oplus {\bf{\bar{\cal{A}}}}_1 \oplus {\bf{\bar{\cal{A}}}}_2, $$
The subspaces ${\bf{\cal{A}}}_0 $ and ${\bf{\bar{\cal{A}}}}_0$ form zero-graded subalgebras.
These algebras can be made {\it unital} if we add to each of them the unit element $\bf{1}$ acting as identity
and considered as being of grade $0$.

If we want the products between the generators $\theta^A$ and the conjugate ones ${\bar{\theta}}^{\dot{B}}$
to be included into the greater algebra spanned by both types of generators, we should
consider all possible products, which will be included in the linear subspaces with a definite grade.
of the resulting algebra ${\cal{A}} \otimes {\bar{\cal{A}}}$.
In order to decide which expressions are linearly dependent,
and what is the overall dimension of the enlarged algebra generated by $\theta^A$'s and
their conjugate variables ${\bar{\theta}}^{\dot{D}}$'s, we must impose on them some
binary commutation relations.

The fact that the conjugate generators are of grade $2$ may suggest that
they behave like products of two ordinary generators $\theta^A \theta^B$. 
Such a choice was often made (see, e.g., \cite{Kerner3}, \cite{Kerner5} and \cite{VARKBLR}).
However, this does not enable one to make a distinction between the
conjugate generators and the products of two ordinary ones, and it would be better
to be able to make the difference. Due to the binary nature of ``mixed"  products, another
choice is possible, namely, to impose the following relations:
\begin{equation}
\theta^{A} {\bar{\theta}}^{\dot{B}} = - j \, {\bar{\theta}}^{\dot{B}} \theta^{A}, \, \ \ \, \ \ 
{\bar{\theta}}^{\dot{B}} \theta^{A} = - j^2 \,\theta^{A} {\bar{\theta}}^{\dot{B}},
\label{commutation2}
\end{equation}
In what follows, we shall deal with the first two simplest realizations of such algebras, 
spanned by two or three generators. Consider the case when $A, B,.. = 1,2$.
The algebra {\cal{A}}
contains numbers, two generators of grade $1$, $\theta^1$ and $\theta^2$, their four
independent products (of grade $2$), and two independent cubic expressions, 
$\theta^1 \theta^2 \theta^1$ and $\theta^2 \theta^1 \theta^2$. Similar expressions can
be produced with conjugate generators ${\bar{\theta}}^{\dot{C}}$; finally, mixed expressions
appear, like four  independent grade $0$ terms $\theta^1 {\bar{\theta}}^{\dot{1}}$,
$\theta^1 {\bar{\theta}}^{\dot{2}}$, $\theta^2 {\bar{\theta}}^{\dot{1}}$ and 
$\theta^2 {\bar{\theta}}^{\dot{2}}$. 

4. Let us consider multilinear forms defined on the algebra ${\cal{A}} \otimes {\bar{\cal{A}}}$.
Because only cubic relations are imposed on products in ${\cal{A}}$ and in $\bar{\cal{A}}$,
and the binary relations on the products of ordinary and conjugate elements, we shall fix
our attention on tri-linear and bi-linear forms, conceived as mappings of ${\cal{A}} \otimes {\bar{\cal{A}}}$
into certain linear spaces over complex numbers.

Let us consider a tri-linear form $\rho^{\alpha}_{ABC}$. Obviously, as
$$\rho^{\alpha}_{ABC} \, \theta^A  \theta^B  \theta^C = \rho^{\alpha}_{BCA} \,
\theta^B  \theta^C  \theta^A = \rho^{\alpha}_{CAB} \, \theta^C  \theta^A  \theta^B, $$
by virtue of the commutation relations (\ref{ternary1}) it follows that we must have
\begin{equation}
\rho^{\alpha}_{ABC} = j^2 \, \rho^{\alpha}_{BCA} = j \,  \rho^{\alpha}_{CAB}.
\label{defrhomatrix3}
\end{equation}
Even in this minimal and discrete case, there are covariant and contravariant
indices: the lower and the upper indices display the inverse transformation property. If a given
cyclic permutation is represented by a multiplication by $j$ for the upper indices, the same permutation performed 
on the lower indices is represented by multiplication by the inverse, i.e. $j^2$, so that they compensate each other.
Similar reasoning leads to the definition of the conjugate forms $ \rho^{{\dot{\alpha}}}_{{\dot{C}}{\dot{B}}{\dot{A}}}$
satisfying the relations (\ref{defrhomatrix3}) with $j$ replaced by $j^2$:
\begin{equation}
{\bar{\rho}}^{{\dot{\alpha}}}_{{\dot{A}}{\dot{B}}{\dot{C}}} = 
j^2  {\bar{\rho}}^{{\dot{\alpha}}}_{{\dot{B}}{\dot{C}}{\dot{A}}}
= j  {\bar{\rho}}^{{\dot{\alpha}}}_{{\dot{C}}{\dot{A}}{\dot{B}}}
\label{defrhomatrix4}
\end{equation}
In the case of two generators, there are only two independent sets of indices. Therefore the
upper indices $\alpha, {\dot{\beta}}$ take on the values $1$ or $2$. We choose the following notation:
\begin{equation}
\rho^1_{121} = j \rho^1_{112} = j^2 \rho^1_{211}; \, \ \ \, \rho^2_{212} = j \rho^2_{221} = j^2 \rho^2_{122},
\label{rhodefinition}
\end{equation}
all other components identically vanishing. The conjugate matrices 
${\bar{\rho}}^{\dot{\alpha}}_{{\dot{B}}{\dot{C}}{\dot{A}}}$ are defined by the same formulae, with $j$
replaced by $j^2$ and vice versa.

5. The constitutive cubic relations between the generators of the $Z_3$ graded algebra
can be considered as intrinsic if they are conserved after linear transformations with commuting 
(pure number) coefficients, i.e. if they are independent of the choice of the basis.
Let $U^{A'}_A$ denote a non-singular $N \times N$ matrix, transforming the generators
$\theta^A$ into another set of generators, $\theta^{B'} = U^{B'}_B \, \theta^B$.
The primed indices run over the same range of values, i.e. from $1$ to $2$; the prime is there
just to make clear we are referring to a new basis.

We are looking for the solution of the covariance condition for the $\rho$-matrices:
\begin{equation}
\Lambda^{{\alpha}'}_{\beta} \, \rho^{{\beta}}_{ABC} = U^{A'}_{A} \, U^{B'}_B \, U^{C'}_C \, 
\rho^{{\alpha}'}_{A' B' C'}.
\label{covtrans1}
\end{equation}
Let us write down the explicit expression, with fixed indices $(ABC)$ on the left-hand side.
Let us choose one of the two available combinations of indices, $(ABC) = (121)$; then the upper
index of the $\rho$-matrix is also fixed and equal to $1$:
\begin{equation}
\Lambda^{{\alpha}'}_{1} \, \rho^{1}_{121} = U^{A'}_{1} \, U^{B'}_2 \, U^{C'}_1 \, 
\rho^{{\alpha}'}_{A' B' C'}.
\label{invariant0}
\end{equation}
Now, $\rho^{1}_{121} = 1$, and we have two equations corresponding to the choice of values of the index $\alpha'$
equal to $1$ or $2$. For $\alpha' = 1'$ the $\rho$-matrix on the right-hand side is $\rho^{1'}_{A' B' C'}$,
which has only three components,
$$\rho^{1'}_{1' 2' 1'}=1, \, \ \ \, \rho^{1'}_{2' 1' 1'}=j^2, \, \ \ \, \rho^{1'}_{1' 1' 2'}=j, $$
which leads to the following equation:
$$\Lambda^{1'}_{1} = U^{1'}_{1} \, U^{2'}_2 \, U^{1'}_1 + j^2 \, U^{2'}_{1} \, U^{1'}_2 \, U^{1'}_1 
+ j \, U^{1'}_{1} \, U^{1'}_2 \, U^{2'}_1 =$$
\begin{equation}
= U^{1'}_{1} \, (U^{2'}_2 \, U^{1'}_1 - U^{2'}_{1} \, U^{1'}_2) = U^{1'}_{1} \, [det(U)],
\label{invariant1}
\end{equation} 
because $j^2 + j = - 1$.
For the alternative choice $\alpha' = 2'$ the $\rho$-matrix on the right-hand side is $\rho^{2'}_{A' B' C'}$,
whose three non-vanishing components are
$$\rho^{2'}_{2' 1' 2'}=1, \, \ \ \, \rho^{2'}_{1' 2' 2'}=j^2, \, \ \ \, \rho^{2'}_{2' 2' 1'}=j. $$
the corresponding equation gives:
\begin{equation}
\Lambda^{2'}_{1} = - U^{2'}_{1} \, [det(U)],
\label{invariant2}
\end{equation} 
The remaining two equations are obtained in a similar manner, resulting in the following: 
\begin{equation}
\Lambda^{1'}_{2} = - U^{1'}_{2} \, [det(U)], \, \ \ \, \ \ \Lambda^{2'}_{2} = U^{2'}_{2} \, [det(U)].
\label{invariant34}
\end{equation}
The determinant of the $2 \times 2$ complex matrix $U^{A'}_B$ appears everywhere on the right-hand side.
Taking the determinant of the matrix $\Lambda^{{\alpha}'}_{\beta}$ one gets immediately
\begin{equation}
det \, ( \Lambda ) = [ det \, (U) ]^3.
\label{detLambdaU}
\end{equation}
Taking into account that the inverse transformation should exist and have the same properties,
we arrive at the conclusion that $det \, \Lambda = 1$,
\begin{equation}
det \, (\Lambda^{\alpha'}_{\beta} ) = \Lambda^{1'}_1 \, \Lambda^{2'}_2 - \Lambda^{2'}_{1} \, \Lambda^{1'}_2 = 1
\label{detlambda}
\end{equation}
which defines the $SL(2, {\bf C})$ group, the covering group of the Lorentz group.

However, the $U$-matrices on the right-hand side are defined only up to the phase, which due to the
cubic character of the relations (\ref{invariant1} - \ref{invariant34}), and they can take on three different values:
$1$, $j$ or $j^2$, i.e. the matrices $j \, U^{A'}_B$ or $j^2 \,  U^{A'}_B$ satisfy the same relations
as the matrices $U^{A'}_B$ defined above. 
The determinant of $U$ can take on the values $1, \, j \,$ or $j^2$ while $det (\Lambda) = 1$

Let us then choose the matrices $\Lambda^{\alpha'}_{\beta}$ to be the usual spinor representation of
the $SL(2, {\bf C})$ group, while the matrices $U^{A'}_{B}$ will be defined as follows:
\begin{equation}
U^{1'}_{1} = j \Lambda^{1'}_1,   U^{1'}_{2} = - j \Lambda^{1'}_2, 
U^{2'}_{1} = - j  \Lambda^{2'}_1,  U^{2'}_{2} = j \Lambda^{2'}_2, 
\label{Umatrices}
\end{equation}
the determinant of $U$ being equal to $j^2$. 

Obviously, the same reasoning leads to the conjugate cubic representation of $SL(2, {\bf C})$ if we require
the covariance of the conjugate tensor 
$${\bar{\rho}}^{\dot{\beta}}_{{\dot{D}}{\dot{E}}{\dot{F}}} = j \, 
{\bar{\rho}}^{\dot{\beta}}_{{\dot{E}}{\dot{F}} {\dot{D}}}
= j^2 \, {\bar{\rho}}^{\dot{\beta}}_{{\dot{F}} {\dot{D}} {\dot{E}}},$$
by imposing the equation similar to (\ref{covtrans1})
\begin{equation}
\Lambda^{{\dot{\alpha}}'}_{{\dot{\beta}}} \, {\bar{\rho}}^{{\dot{\beta}}}_{{\dot{A}}{\dot{B}}{\dot{C}}} = 
{\bar{\rho}}^{{\dot{\alpha}}'}_{{\dot{A}}' {\dot{B}}' {\dot{C}}'} {\bar{U}}^{{\dot{A}}'}_{{\dot{A}}} \, 
{\bar{U}}^{{\dot{B}}'}_{{\dot{B}}} \, {\bar{U}}^{{\dot{C}}'}_{{\dot{C}}} .
\label{covtrans2}
\end{equation}
The matrix $\bar{U}$ is the complex conjugate of the matrix $U$, and its determinant is equal to $j$.

Moreover, the two-component entities obtained as images of cubic combinations of quarks,
$\psi^{\alpha} = \rho^{\alpha}_{ABC} \theta^A \theta^B \theta^C$ 
 and ${\bar{\psi}}^{\dot{\beta}} = {\bar{\rho}}^{{\dot{\beta}}}_{{\dot{D}}{\dot{E}}{\dot{F}}}
{\bar{\theta}}^{\dot{D}} {\bar{\theta}}^{\dot{E}} {\bar{\theta}}^{\dot{F}} $ should anti-commute,
because their arguments do so, by virtue of (\ref{commutation2}):
$$ (\theta^A \theta^B \theta^C) ({\bar{\theta}}^{\dot{D}} {\bar{\theta}}^{\dot{E}} {\bar{\theta}}^{\dot{F}} )
= - ({\bar{\theta}}^{\dot{D}} {\bar{\theta}}^{\dot{E}} {\bar{\theta}}^{\dot{F}})(\theta^A \theta^B \theta^C)$$

We have found the way to derive the covering group of the Lorentz group acting on spinors via
the usual spinorial representation. The spinors are obtained as the homomorphic image of 
tri-linear combination of three quarks (or anti-quarks). The quarks transform with
matrices $U$ (or ${\bar{U}}$ for the anti-quarks), but these matrices are not unitary: 
their determinants are equal to $j^2$ or $j$, respectively. So, quarks cannot
be put on the same footing as classical spinors; they transform under a $Z_3$-covering of
the $SL(2, {\bf C})$ group.

A similar covariance requirement can be formulated with respect to the set of $2$-forms
mapping the quadratic quark-anti-quark combinations into a four-dimensional linear real space.
As we already saw, the symmetry (\ref{commutation2}) imposed on these expressions reduces their
number to four. Let us define two quadratic forms, $\pi{\mu}_{A {\dot{B}}}$ and its conjugate
${\bar{\pi}}^{\mu}_{{\dot{B}} A}$ with the following symmetry requirement
\begin{equation}
\pi^{\mu}_{A {\dot{B}}} \, \theta^A {\bar{\theta}}^{\dot{B}} = {\bar{\pi}}^{\mu}_{{\dot{B}} A}
\, {\bar{\theta}}^{\dot{B}} \theta^A.
\label{pisymmetry1}
\end{equation}
The Greek indices $\mu, \nu...$ take on four values, and we shall label them
$0,1,2,3$.
It follows immediately from (\ref{commutation2}) that
\begin{equation}
\pi^{\mu}_{A {\dot{B}}} = - j^2 \, {\bar{\pi}}^{\mu}_{{\dot{B}} A}.
\label{pisymmetry2}
\end{equation}
Such matrices are non-hermitian, and they can be realized by the following substitution:
\begin{equation}
\pi^{\mu}_{A {\dot{B}}} = j^2 \, i \, {\sigma}^{\mu}_{A {\dot{B}}}, \, \ \ \,
{\bar{\pi}}^{\mu}_{{\dot{B}} A} = - j \, i \, {\sigma}^{\mu}_{{\dot{B}} A}
\label{pidefinition2}
\end{equation}
where ${\sigma}^{\mu}_{A {\dot{B}}}$
are the unit $2 \time 2$ matrix for $\mu = 0$, and the three hermitian Pauli matrices for $\mu = 1,2,3$.

Again, we want to get the same form of these four matrices in another basis. Knowing
that the lower indices $A$ and ${\dot{B}}$ undergo the transformation with matrices $U^{A'}_B$
and ${\bar{U}}^{{\dot{A}}'}_{\dot{B}}$, we demand that there exist some $4 \times 4$ matrices
$\Lambda^{{\mu}'}_{\nu}$ representing the transformation of lower indices by the matrices
$U$ and ${\bar{U}}$:
\begin{equation}
\Lambda^{{\mu}'}_{\nu} \, \pi^{\nu}_{A {\dot{B}}} = U^{A'}_A \, {\bar{U}}^{{\dot{B}}'}_{\dot{B}}
 \pi^{{\mu}'}_{A' {\dot{B}}'},
\label{pitransform1}
\end{equation}
and this defines the vector ($4 \times 4)$ representation of the Lorentz group.
Introducing the invariant ``spinorial metric" in two complex dimensions, $\varepsilon^{AB}$
and $\varepsilon^{{\dot{A}}{\dot{B}}}$ such that $\varepsilon^{12} = - \varepsilon^{21} = 1$
and $\varepsilon^{{\dot{1}}{\dot{2}}} = - \varepsilon^{{\dot{2}}{\dot{1}}}$, we can define
the contravariant components $\pi^{\nu \, \, A {\dot{B}}}$. It is easy to show that the
Minkowskian space-time metric, invariant under the Lorentz transformations, can be defined as
\begin{equation}
g^{\mu \nu} = \frac{1}{2} \biggl[ \pi^{\mu}_{A {\dot{B}}} \, \pi^{\nu \, \, A {\dot{B}}} \biggr] 
= diag (+,-,-,-)
\label{Mmetric}
\end{equation}
Together with the anti-commuting spinors ${\psi}^{\alpha}$ the four real coefficients defining
a Lorentz vector, $x_{\mu} \, {\pi}^{\mu}_{A {\dot{B}}}$, can generate now the supersymmetry
via standard definitions of super-derivations.

6. Consider now three generators, $Q^a , \, \, a=1,2,3$, and their conjugates
${\bar{Q}}^{\dot{b}}$ satisfying similar cubic commutation relations as in the two-dimensional case:

\vskip 0.2cm
\centerline{$Q^a Q^b Q^c = j \, Q^b Q^c Q^a = j^2 \, Q^c Q^a Q^b,$}
\vskip 0.2cm
\centerline{${\bar{Q}}^{\dot{a}} {\bar{Q}}^{\dot{b}} {\bar{Q}}^{\dot{c}} = 
j^2 \, {\bar{Q}}^{\dot{b}} {\bar{Q}}^{\dot{c}} {\bar{Q}}^{\dot{a}} 
= j \, {\bar{Q}}^{\dot{c}} {\bar{Q}}^{\dot{a}} {\bar{Q}}^{\dot{b}}$,}
\vskip 0.2 cm
\centerline{$Q^a \, {\bar{Q}}^{\dot{b}} = - j {\bar{Q}}^{\dot{b}} \, Q^a.$}
\vskip 0.2cm

With the indices $a, b ,c...$ ranging from $1$ to $3$ we get {\it eight} linearly independent
combinations of three undotted indices, and the same number of combinations of dotted ones.
They can be arranged as follows:
\vskip 0.2cm
\centerline{$Q^3 Q^2 Q^3, \, \ \  Q^2 Q^3 Q^2,  \, \ \  Q^1 Q^2 Q^1, $}
\vskip 0.2cm
\centerline{$ Q^3 Q^1 Q^3,  \, \ \  Q^1 Q^2 Q^1, \, \ \  Q^2 Q^1 Q^2,$}
\vskip 0.2cm
\centerline{$Q^1 Q^2 Q^3, \, \ \ \, Q^3 Q^2 Q^1,$}
\vskip 0.2cm
\noindent
while the quadratic expressions of grade $0$, $Q^a \, {\bar{Q}}^{\dot{b}}$ span
a $9$-dimensional subspace in the finite algebra generaterd by $Q^a$'s.
The invariant 3-form mapping these combinations onto some eight-dimensional space must
have also eight independent components (over real numbers). The 
three-dimensional ``cubic matrices" are then as follows:
\vskip 0.2cm
\centerline{$ K^{3+}_{121} = 1, \, \ \ \, K^{3+}_{112} = j^2, \, \ \ \, K^{3+}_{211} = j;$} 
\vskip 0.2cm
\centerline{$K^{3-}_{212} = 1, \, \ \ \, K^{3-}_{221} = j^2, \,\ \ \,  K^{3-}_{122} = j;$}
\vskip 0.2cm
\centerline{$ K^{2+}_{313} = 1, \, \ \ \, K^{2+}_{331} = j^2, \, \ \ \, K^{2+}_{133} = j;$}
\vskip 0.2cm
\centerline{$K^{2-}_{131} = 1, \, \ \ \,  K^{2-}_{113} = j^2, \, \ \ \,  K^{2-}_{311} = j;$}
\vskip 0.2cm
\centerline{$ K^{1+}_{232} = 1, \, \ \ \, K^{1+}_{223} = j^2, \, \ \ \, K^{1+}_{322} = j;$} 
\vskip 0.2cm
\centerline{$K^{1-}_{323} = 1, \, \ \ \, K^{1-}_{332} = j^2, \, \ \ \, K^{1-}_{233} = j;$}
\vskip 0.2cm
\centerline{$K^7_{123} = 1, \, \ \ \, K^7_{231} = j^2, \, \ \ \,  K^7_{312} = j;$} 
\vskip 0.2cm
\centerline{$K^8_{132} = 1, \, \ \ \, K^8_{321} = j^2, \ \ \,  \, K^8_{213} = j.$}
\vskip 0.2cm
\noindent
all other components being identically zero.
Let the capital Greek indices $ \Phi, \Omega$ take on the values from $1$ to $8$. The
covariance principle applied to the cubic matrices $K^{\Phi}_{abc}$ underlinear change
 of the basis from $\theta^a$ to $\theta{a'}= U^{a'}_b \, \theta^b$ means that we want
to solve the following equations:
\begin{equation}
S^{{\Phi}'}_{\Omega} \, K^{\Omega}_{d e f} = K^{{\Phi}'}_{a' b' c' }
 U^{a'}_d \,  U^{b'}_e \, U^{c'}_f ,
\label{Kmatrixtransf}
\end{equation}
\noindent
It takes more time to prove, but the result is that the $8 \times 8$ matrices $S^{{\Phi}'}_{\Omega}$
are the adjoint representation of the $SU(3)$ group, whereas the $3 \times 3$ matrices $U^{a'}_d$
are the fundamental representation of the same group, up to the phase factor that can take on the
values $1, \, j$ or $j^2$. 

The nine independent two-forms $P^{i}_{a {\dot{b}}} = - j^2 \, {\bar{P}}^{i}_{{\dot{b}} a}$
transform as the $3 \otimes {\bar{3}}$ representation of $SU(3)$

Finally, the elements of the tensor product of both types of $j$-anti-commuting entities,
$\theta^A$ and $Q^b$ can be formed, giving six quarks, $Q^{B}_a$, transforming via
$Z_3$ coverings of $SL(2, {\bf C})$ and $SU(3)$, which looks very much like the three flavors.

7. We have shown how the requirement of covariance of $Z_3$-graded
cubic generalization of anti-commutation relations leads to spinor and vector representations
of the Lorentz group and the fundamental and adjoint representations of the $SU(3)$ group, 
thus giving the cubic $Z_3$-graded quark algebra primary role in determining the
Lorentz and $SU(3)$ symmetries. However, these representations coincide with the usual ones
only when applied to special combinations of quark variables, cubic (spinor)
or quadratic (vector) representations of the Lorentz group. 

While acting on quark variables, the representations correspond to the $Z_3$-covering of 
groups. In this sense quarks are not like ordinary spinors or fermions, and as such, 
do not obey the usual Dirac equation. If the sigma-matrices are to be replaced by the
non-hermitian matrices $\pi^{\mu}_{A {\dot{B}}}$, instead of the usual wave-like solutions
of Dirac's equation we shall get the exponentials of complex wave vectors, and such solutions
cannot propagate. Nevertheless, as argued in \cite{Kerner5}, certain tri-linear and bi-linear combinations
of such solutions behave as usual plane waves, with real wave vectors and frequencies,
if there is a convenient coupling of non-propagating solutions in the 
$k$-space.
\vskip 0.2cm
\indent 
{\bf Acknowledgment.}
We are greatly indebted to Michel Dubois-Violette for numerous discussions and enlightening remarks.

\end{document}